\documentclass[review]{elsarticle}

\usepackage{lineno,hyperref}
\usepackage{algpseudocode}
\usepackage[english]{babel}
\usepackage[utf8]{inputenc}
\usepackage{algorithm}
\usepackage{ifthen}

\usepackage{graphicx}
\usepackage{multirow}
\graphicspath{ {figures/} }

\begin{document}

\begin{frontmatter}

\title{A Simulation Study of Social-Networking-Driven Smart Recommendations for IoV}

\author{Kashif Zia, Arshad Muhammad, Dinesh Kumar Saini}
\address{Sohar University, Sohar, Oman}

\begin{abstract}
Social aspects of connectivity and information dispersion are often ignored while weighing the potential of Internet of Things (IoT). In the specialized domain of Internet of Vehicles (IoV), Social IoV (SIoV) is introduced realization its importance. Assuming a more commonly acceptable standardization of Big Data generated by IoV, the social dimensions enabling its fruitful usage remains a challenge. In this paper, an agent-based model of information sharing between vehicles for context-aware recommendations is presented. The model adheres to social dimensions as that of human society. Some important hypotheses are tested under reasonable connectivity and data constraints. The simulation results reveal that closure of social ties and its timing impacts dispersion of novel information (necessary for a recommender system) substantially. It was also observed that as the network evolves as a result of incremental interactions, recommendations guaranteeing a fair distribution of vehicles across equally good competitors is not possible.  
\end{abstract}

\begin{keyword}
Social IoV \sep Agent-Based Model \sep Discrete-Event Simulation\sep SIoT
\end{keyword}

\end{frontmatter}


\section{Introduction}
Modern technology has reached to a point whereby electronic devices (handheld, wearable, sensors, appliances etc.) are commonplace in every facet of our life. In 1991, Mark Weiser \cite{weiser1991computer} visioned it and described that in the 21st century, computers will become so ubiquitous that they pervade every area of our lives and environment. He believed the development will be inevitable, and that the effects would be far greater than just the physical provisioning of devices. He believed that the technology would become a fabric of everyday life, changing the way we interact with our surrounding \cite{gubbi2013internet}, including both the environment and the society. 

The latest manifestation of ``all connected world'' is Internet of Things (IoT) \cite{wortmann2015internet}. At the physical level, objects in IoT have embedded processors and capability to communicate among them using wired or wireless connections \cite{bakici2013smart,neirotti2014current}. According to the recent predictions by CISCO, 50 billion ``things'' will be connect to the Internet by 2020 \cite{evans2011internet}. Vehicles being an important part of our lives and first one to be equipped with recent technologies, are rapidly becoming the first case study of IoT, more appropriately termed as Internet of Vehicles (IoV) \cite {fangchun2014overview, dandala2017internet}.

IoV is more than numerous sensors embedded in modern vehicles which can not only receive the information from its surroundings but also transmit information, assisting in navigation and traffic management \cite{alam2015toward}. A practical realization of vehicular technologies is  Vehicular Ad hoc NETworks (VANETs) \cite{cunha2016data}, which resulted in important applications regarding handling traffic and better driver / passengers experience \cite{alam2015toward}. On the other side, we have Vehicular Social Networks (VSN) where passengers can exchange data related to entertainment, social networks and situations \cite{abbani2011managing} However, IoV is more then a social network of vehicles themselves; in which, vehicles build and manage their own social network; more appropriately termed as Social Internet of Vehicles (SIoV).

SIoV is a social network of vehicles, still, it cannot be separated from the drivers or owners of these vehicles. Hence, when a vehicle is a part of SIoV, it must build and use the network to achieve owners' goals. Although, it can be debated, but the properties of human social network are mapped onto buildup and evolution of inter-vehicle social network. Then, such a ``physical'' SIoV is used as a medium to achieve owners' goals. The interfacing of the network and the extend of user (owner of vehicle) satisfaction with respect to goal achievement is a gray area for the purpose of this paper, but definitely possible using current personal, vehicular and domestic technologies.  

This paper investigates the potential of SIoV by designing an agent-based model and simulating in various conditions. The model's conceptualization is based on basic assumptions about IoT connectivity, principles of social network evolution, and users profiling. User profiles and activities serve as an underpinning for the model. Some interesting what-if questions are asked against a couple of intuitive hypotheses. However, the study revolves around finding the extend to which SIoV is capable of providing correct recommendations to their owners in the wake of changing dynamics of the resources. 

The rest of the paper is organized as follows. Related work follows in section \ref{sec:rw}. In section \ref {sec:hyp}, we present the scenario adopted and research hypotheses adapted from the scenario.  The proposal model is explained in section \ref{sec:model}. Simulation settings and analysis of the simulation results is presented in section \ref {sec:sim}. Section \ref {sec:disc} provides discussion about the hypothesis and conclusion. 

\section{Related Work} \label {sec:rw}

Recently there are several research activities to investigate the possibilities of embedding social networking concepts into the Internet of Things (IoT) solutions; resulting paradigm is known as Social Internet of Things (SIoT) \cite{atzori2012social}. It is evident from the research that a group of people or society (sharing the same interest) can offer more accurate information as compared to an individual. In \cite{atzori2015social}, define SIoT as IoT where things are capable of establishing the social relationship with other objects autonomously. Social networking concepts have been applied in several networking settings such as delay tolerant, peer-to-peer networking \cite{atzori2014smart}. Authors in \cite{atzori2014smart} suggest that objects can establish social relationships based on object profile, activities, and interests.

In \cite{alam2015toward} introduce an exciting idea of Social Internet of Vehicle (SIoV) using existing VANET’s technologies such as vehicle-to-vehicle, vehicle-to-infrastructure and, vehicle-to-internet communications and presents a vehicular social network platform following cyber-physical \cite{lee2015cyber}. SIoV uses social relationships between physical objects to exchange and stores different types of information such as safety, infotainment, comfort etc. as a social graph. It provides online or offline information for intelligent transport system \cite{festag2014cooperative}, online allows safe and efficient travel of the vehicles while offline data ensures smart behavior of the vehicles.

\cite{smaldone2008roadspeak} presented an exciting idea of Vehicular Social Networks (VSNs) that a large number of people spend 2-3 hours daily on the road to commute between home and the office. These users use the same roads on a daily basis provides an excellent opportunity to form periodic virtual mobile communities. Three primary reasons for VSN formation are for entertainment (based on people's interest to discuss music, movies etc), utility(local event's and/or Point of Interest (POI) such as hotel, shopping center)  and emergency (road accidents or assistance during critical conditions). Authors presented RoadSpeak a VSN-based system allows driver to join Voice Chat Groups (VCGs) along the popular highways and roadways to facilitate communication among the car occupants. 

\cite{Lei2016} presented a model for optimal route choice to helps drivers to decrease travel time and reduce the traffic congestion \cite{dias2014cooperation} by utilizing a Vehicle agent-based in Edge and cloud to allow drivers to negotiate with others known as Virtual Vehicle (VV) \cite{fangchun2014overview}. Each driver has corresponding VV, and parts of driver's knowledge, which can replace the driver to make a decision in cloud. Using bargaining routing approach for the optimal route calculation. They set up a source node, target nodes and m parallel links, known to all players. The players decided to take the route and realize the cost incur on their route. Maybe due to the selfish behavior of the VV not cooperate with others and in case of cooperation with others to decrease the cost of travel.  

In \cite{hu2013social} presented a Social Drive (crowdsourcing-based VSN) system for green transportation by integrating vehicular On-Board Diagnostics (OBD) module, cloud computing, and social networks and incorporates rating mechanism about the driver's fuel economy. Using a mobile application to promote awareness of their driving behaviors regarding fuel economy. There are numbers of other crowdsourcing based applications for green transportation such as UbiGreen \cite{froehlich2009ubigreen}, GreenGPS \cite{saremi2016participatory}, and Cyberphysical bike\cite{smaldone2011cyber}.

According to predication in \cite{sun2016internet} by 2050, 70\% of world population will live in the cities. The concept of smart cities paving the way for Smart and Connected Communities (SCC) defined as a group of people who are living in the same area such as city or town. Authors argue that IoT can provide a ubiquitous network of connected devices and smart sensors for SCC and big data analytics \cite{zakir2015big} has potential to make a move from IoT to real-time control for SSC. Authors presented TreSight as a case study for smart tourism in the city of Trento, Italy, a context-aware recommendation system to provide personalized recommendations. Using data from OpenData Trentino regarding POI, weather, etc. with additional data collected using CrowdSensing with a wearable bracelet. 

In this paper, we incooperate social relationship among the objects (vehicle refer as a object), i.e. social relationship may be established between these vehicles while there owner's visiting the same PoI. At the start, relationship may be a ‘weak tie’ later converted to ‘strong tie’ depends on the number of times these objects meet each other. Generating user profiles based on the recommendations vehicles may receive from others in the social network. User weekly plan may be changed as per these recommendations by providing/suggesting an alternative PoI, which could be higher or equal expectation of the user. From the above given scenario, an agent-based model is designed. 

\section {Scenario and Research Hypotheses} \label{sec:hyp}

We usually spend most of our time during the weekdays to commute between work, home, school, and shopping. Drivers use number of applications such as navigators \cite{park2014driver} to reduce the travel time or to choose between different available options. Moreover, in recent years, Location Based Social Networks (LBSN) \cite {bao2015recommendations}, i.e. Facebook Places, Googlemaps etc have gained popularity, where users can share their physical locations, experiences, and ratings etc. Due to these developments, recommender or recommendation systems have gained popularity in recent years where big data is driving force to provide context-aware recommendations to their users. 

Realization of a recommendation system enabled by the capabilities of SIoV is an exciting topic. It becomes achievable due to recent industrial endeavors. For example, In CarStream project \cite{zhang2017carstream}, a technological integration is performed to provide data-driven services. Over 30,000 chauffeured driven cars are connected, the system collects a variety of data such as ``vehicle status, driver activity, and passenger-trip information''. The user demand is generated by combining parameters such as pickup point, pickup time, arrive time and destination. Motivated from these data fields, our model scenario utilizes the concepts of \textit{ID}, \textit{Time} and \textit{Duration}, to describe a Point of Interest (PoI) from vehicle or user perspective, where \textit{ID} is the identity of the destination, \textit{Time} is the time to reach the destination, and \textit{Duration} is the time to stay at the destination. 

User profiling (motivated from \cite{fagnant2014travel}) is used to generate a \textit{plan} in which a user has to visit some PoIs on a weekly basis. Each user has an \textit {expectation}, which is a static personal trait. If the \textit{quality} of service provided by a PoI at the time of a visit is not up to the user's expectation (user \textit{experience}), the PoI is considered as \textit {suspended} for that user. Vehicles connect with each other if they are at the same PoI, thus evolving a social network of vehicles with repeated encounters. The question asked in this paper is: to what extent SIoV is capable of providing the required PoIs recommendations to the users if a user has most of PoIs suspended due to poor quality and/or high expectation.

Before, stating the exact hypotheses, it is worth noting that the vehicles are not different from users. The interface between a vehicle and its owner (user) is assumed to be seamless so that a vehicle know about its owner's plan and also if the current experience about a PoI was good or bad. Vehicle is also capable of changing owner's plan without owner approval. To avoid unnecessary complexity, all these simplifications are made to keep our focus on the evaluation of the following research hypotheses: 

\begin{itemize}
\item \textit {Hypothesis 1:} Since connectivity potentially transforms into interaction (information sharing), an increased connectivity degree provides more prospect of having novel information; in short, a \textit{novelty in information is supported by triadic closure} \cite{easley2010networks}.
\item \textit {Hypothesis 2:} Since information sharing should increase the user experience, early the information sharing starts, the better would be the user experience; in short, \textit{novel information sharing is directly proportional to how early the information sharing starts}.
\item \textit {Hypothesis 3:} Due to intrinsic random nature of individual plans, \textit {a fair distribution of resources should be guaranteed as a result of novel information sharing}. 
\end{itemize}

An agent-based model and simulation is used to verify these hypotheses.

\section{Model} \label {sec:model}

The basic purpose of the vehicles is to visit PoIs given in the plan at their prescribed time. The three strategies adopted are:

\begin {enumerate}
\item AsPlanned
\item Blacklist
\item Replace (without and with triadic closure)
\end {enumerate}

The description (section \ref{subsec:strategies}) of these strategies follows section \ref {subsec:flow}.

\subsection {Model Flow} \label {subsec:flow}

\paragraph {Vehicle / User Plan} Plan is a matrix of five columns. The first column contains the $ID$ of a PoI, second column is the $time$ to visit that PoI. The third column is the $duration$ of the visit. The fourth column stores numeric representation of last visit $experience$ of the user (vehicle) about that PoI. The fifth column represents if this PoI is $suspended$ or not. A list of random PoIs are chosen along with corresponding visit times and visit durations, which are also random. One simulation iteration is equal to one hour. Initially, a random weekly plan, constituted by three components (scheduled visits) is generated, which executes on weekly basis.  

\begin{figure}
  \includegraphics[width=\textwidth]{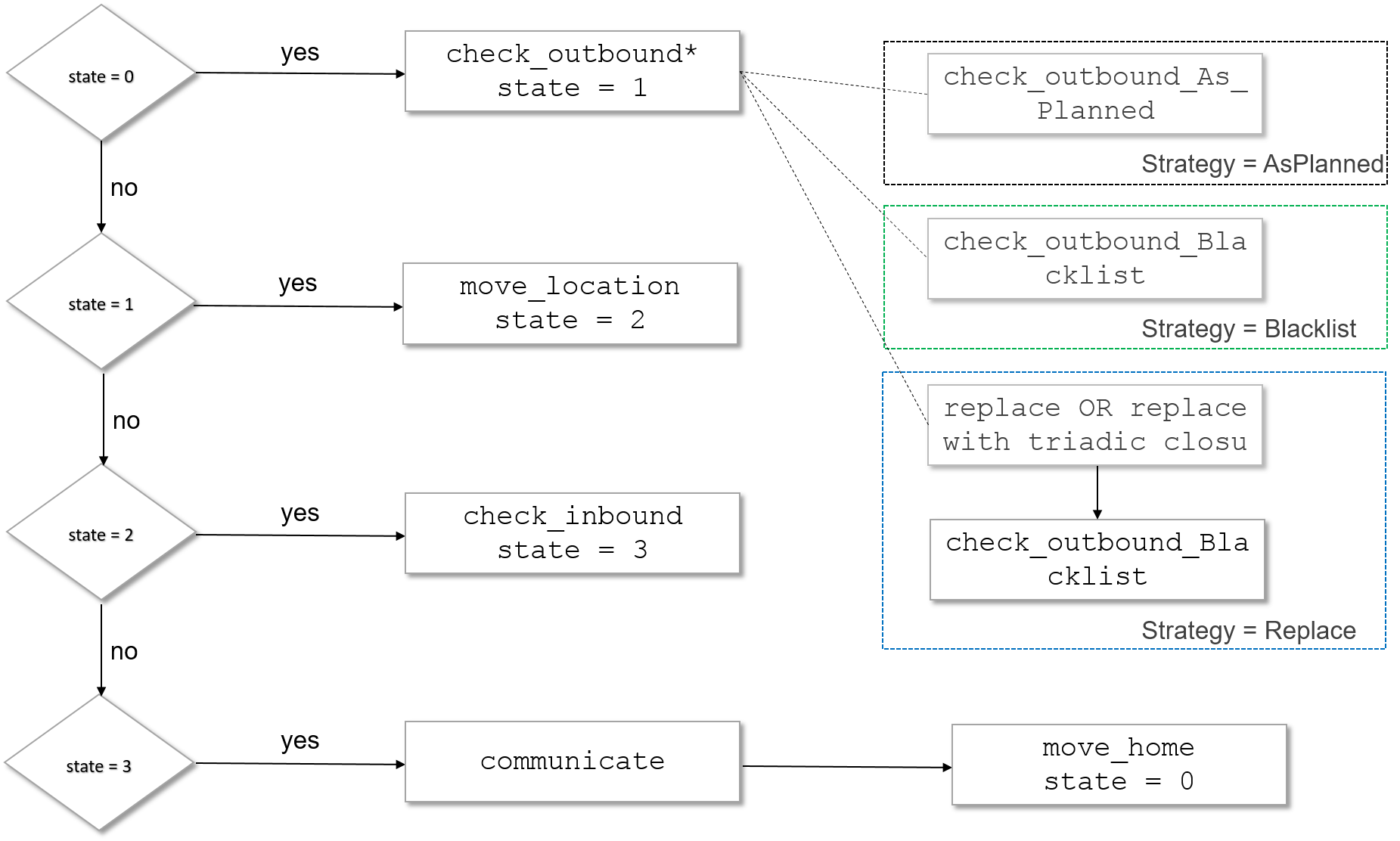}
  \caption{Vehicle state transitions to be performed at each iteration (= one hour) of the simulation by all the vehicles .}
  \label{fig:system1}
\end{figure}

\paragraph{Vehicle States} The transition system of vehicles' behavior is divided into five states. If $state = 0$, the vehicle is located at user's home. A vehicle will transit to the next state ($state = 1$) if it is $time$ to execute a component of the plan. This is done through \texttt{check\_outbound*} procedure. If $state = 1$, the vehicle moves to the desired location along with transiting to the next state ($state = 2$). This is done through \texttt{move\_location} procedure. If $state = 2$, and the $duration$ to stay at the current location is expired, the vehicle will transit to the next state ($state = 3$). This is done through \texttt{check\_inbound} procedure. If $state = 3$, the vehicle will communicate with its neighbors (in proximity). This is done through \texttt{communicate} procedure. After communicating, the vehicle will move back to home and resets its state to $0$. This is done through \texttt{move\_home} procedure. Figure \ref{fig:system1} presents a graphical view of vehicles' state transitions. In the following, a description of each of these procedures is given.     

\begin{table}
  \centering
    \begin{tabular}{ | l | l | l | l | l |}
    \hline
    ID & Time & Duration & Experience & Suspended? \\ \hline
    1059 & 17 & 2 & 1 & false \\ \hline
        1054 & 122 & 4 & 1 & false \\ \hline
            1053 & 50 & 2 & 1 & false \\ \hline
    \end{tabular}
    \caption{Plan matrix Initialization (vehicle 569).}
    \label{tbl:matrixstart}
\end{table}
 
\paragraph{\texttt{check\_outbound*}} The plan matrix contains three components (scheduled visits), which are randomly set at the start of the simulation. Table \ref {tbl:matrixstart} shows an example of this matrix for a sample vehicle, with ID $569$. At hour $17$, this vehicle has to visit PoI $1059$ for a duration equal to $2$ hours. The \texttt{check\_outbound*} procedure, at iteration $17$ would identify it and the state of the vehicle would change from $0$ to $1$. Similarly, at their turn, PoI $1054$ and $1053$ will also be visited. By default, the experience of the user regarding the recent visit to a PoI is set to default positive extreme equal to $1$. As shown in Figure \ref{fig:system1}, the procedure \texttt{check\_outbound*} has three varieties  which correspond to three different strategies of selection of a PoI. Here, we have explained strategy of selecting a PoI according to the initial plan, named as \texttt{check\_outbound\_As\_Planned} (replacing * by the strategy used). Obviously, this strategy is static, just executing the plan as it is. 

\begin{figure}
\centering
  \includegraphics[width=0.75\textwidth]{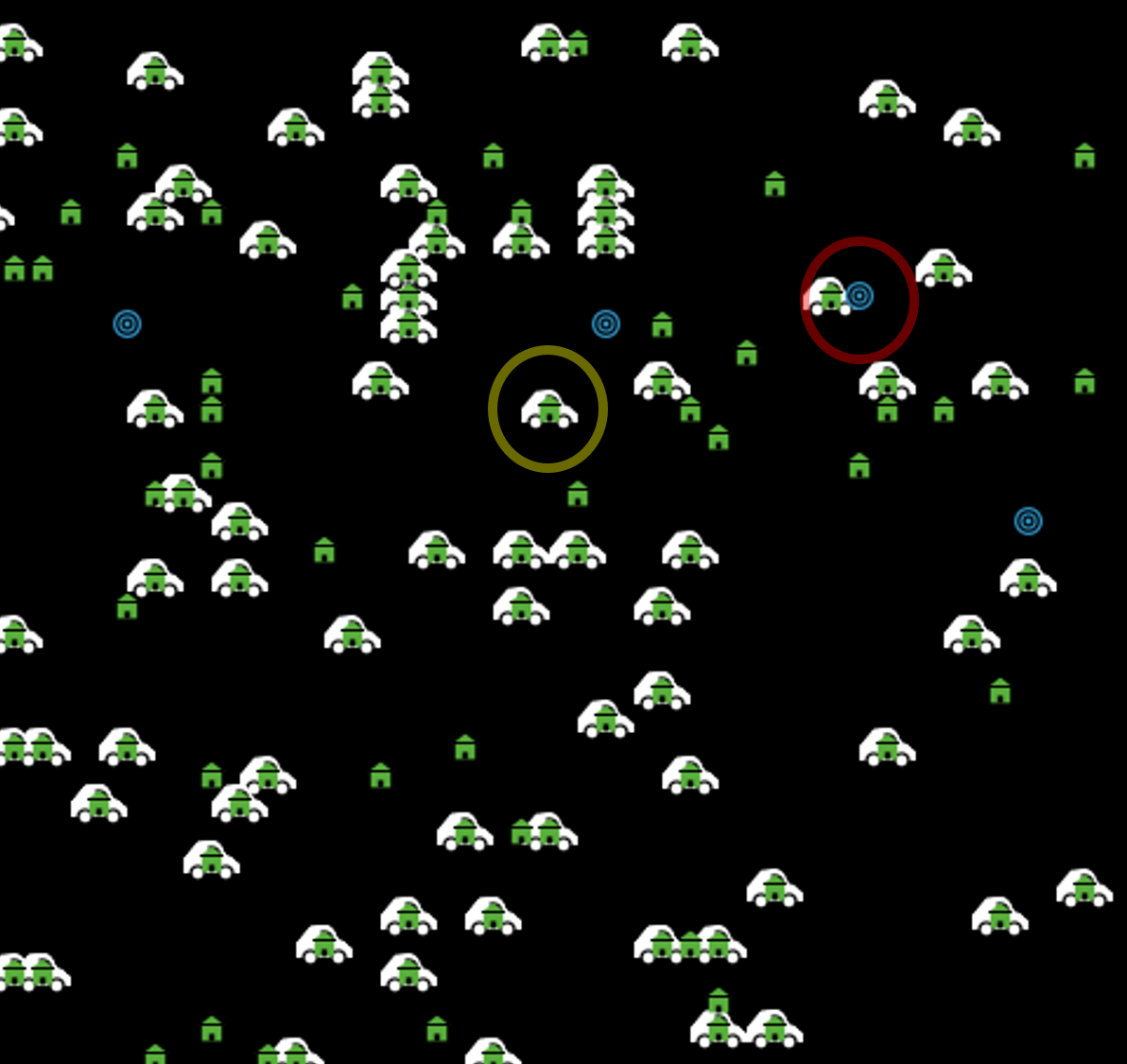}
  \caption{A view of one fourth of the simulation world. Vehicle with yellow circle is vehicle $569$, and PoI with red circle is PoI $1059$.}
  \label{fig:setup}
\end{figure}

\begin{figure}
\centering
  \includegraphics[width=0.75\textwidth]{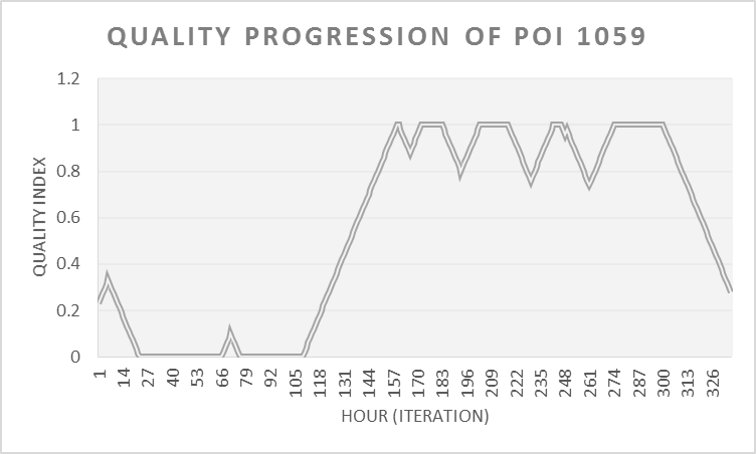}
  \caption{Quality progression of PoI 1059 of first two weeks.}
  \label{fig:1059}
\end{figure}

\paragraph{\texttt{move\_location}} A vehicle moves to the PoI selected in \texttt{check\_outbound*} procedure. The current $quality$ offered by the PoI with a random variation of $\pm 25$ \% is stored in the plan matrix as experience of the user. For example, vehicle $569$ has to visit PoI $1059$ at iteration 17 (also see Figure \ref{fig:setup}). The progression curve of $quality$ offered by PoI $1059$ is shown in \ref {fig:1059}. At iteration 17, the value is $0.092$ which is transformed into $experience$ of the vehicle and stored into plan matrix as $0.0909$ (see updated matrix of Table \ref{tbl:upexp}).   

\begin{table}
  \centering
    \begin{tabular}{ | l | l | l | l | l |}
    \hline
    ID & Time & Duration & Experience & Suspended? \\ \hline
    1059 & 17 & 2 & 0.0909 & false \\ \hline
        1054 & 122 & 4 & 1 & false \\ \hline
            1053 & 50 & 2 & 1 & false \\ \hline
    \end{tabular}
    \caption{Updated plan matrix (vehicle 569).}
    \label{tbl:upexp}
\end{table}

\paragraph{\texttt{check\_inbound}} If visit duration of a vehicle at a PoI is exhausted, the state of the vehicle transits into $3$. After completing 2 units of duration at PoI 1059, vehicle 569's state changes to 3 at iteration 20.  

\paragraph{\texttt{communicate}} The vehicles co-located at the same PoI network with each other to form ties. Initially, it is a weak tie. With repeated encounters, a weak tie becomes a strong tie. Each vehicle has a data structure, which contains this information. For each vehicle which has formed a tie with another vehicle, the array of information is constituted by a tuple ``time of the latest encounter, how many encounters?, strong tie?''. If number of encounters with the same vehicle reaches to a \textit{threshold}, a weak tie changes into a strong tie.

For example, vehicle 259 at iteration 20 encounters two more vehicles at PoI 1059. It is evident from Table \ref {tbl:vd259}, both these vehicles are encountered for the first time (how many encounters? = 1) and have established a weak tie with vehicle 259 (strong tie? = false). With repeated encounters at same or other locations, a weak tie will change into a strong tie, if the number of encounters are equal to the \textit{threshold}. 

\begin{table}
  \centering
\begin{tabular}{ |l|l|l| }
\hline
\multicolumn{3}{ |c| }{Vehicles Data of vehicle 569} \\ \hline
\multirow{3}{*}{984} & time of latest encounter & 20 \\
 & how many encounters? & 1 \\
 & strong tie? & false \\ \hline
\multirow{3}{*}{691} & time of latest encounter & 20 \\
 & how many encounters? & 1 \\
 & strong tie? & false \\ \hline

\end{tabular}
\caption{Vehicles Data (vehicle 569).}
    \label{tbl:vd259}
\end{table}

It is worth noting that communication is not used in first two strategies. Hence, in first two strategies, after completion of duration to stay at a PoI, a vehicle moves back to its home.  

\paragraph{\texttt{move\_home}} A vehicle moves back to its home, resetting its state from 3 to 0.

\subsection {Strategies} \label {subsec:strategies} 

\paragraph{\textbf{AsPlanned}} implemented through \texttt{check\_outbound\_AsPlanned} as explained in the last subsection. 

\begin{algorithm}
\scriptsize
\caption{Check\_outbound\_blacklist Algorithm}\label{alg:checkOB}
\begin{algorithmic}[1]
\State $i \gets $0
\While{($i < 3$)}

\State ${t} \gets $plan[i,1]
\State ${q} \gets $plan[i,3]
\State ${s} \gets $plan[i,4]

\If{$(t = current\_time \&\& s=false)$}            
	\If{$(q > expectation)$} 
     		 \State $state \gets 1$
	      \State $select \ ID \ of \ this \ row \ as \ destination$
     \Else
   	     \State $plan[i,4] \gets true$
	\EndIf
\EndIf
     		\State $j \gets $0
        \While{$j < 3$}
        	\State $qq \gets $plan[j,3]
            \If{$(qq >= expectation)$} 
            	\State $plan[j,1] \gets current\_time $
                \State $state \gets 1$
                \State $select \ ID \ of \ this \ row \ as \ destination$
                \State $j \gets $3
            \Else
            	\State $j \gets $j++
\EndIf
\EndWhile
\EndWhile
\end{algorithmic}
\end{algorithm}

\paragraph{\textbf{Blacklist}} implemented through \texttt{check\_outbound\_Blacklist}. From a vehicle's plan, the method of selection of a visits works as follows. If the vehicle has to visit a PoI at the current hour and that PoI is not already blacklisted (if blacklisted then do not visit any location), then the quality of current PoI will be evaluated against vehicle's $expectation$. If quality is greater than expectation, the vehicle will transit to state 1, ready to visit the current PoI. Otherwise, the PoI will be suspended in the plan matrix. For example, at the 17th hour of the second week, the last experience of vehicle 569 about PoI 1059 was 0.0909, which is well below its expectation (0.465640). Hence, PoI 1059 will be suspended in vehicle 569 plan (see Table \ref{tbl:matw217}). Further, an alternate PoI would be sought that is not yet blacklisted to be executed in the current hour. If no such PoI is found, the vehicle would not visit any location (state remains 0). Algorithm \ref{alg:checkOB} provides pseudocode of the procedure. 

\begin{table}
  \centering
    \begin{tabular}{ | l | l | l | l | l |}
    \hline
    ID & Time & Duration & Experience & Suspended? \\ \hline
    1059 & 17 & 2 & 0.0909 & true \\ \hline
        1054 & 122 & 4 & 0.5684 & false \\ \hline
            1053 & 50 & 2 & 0.3325 & false \\ \hline
    \end{tabular}
    \caption{Plan matrix update at 17th hour of second week (vehicle 569).}
    \label{tbl:matw217}
\end{table}

\paragraph{\textbf{Replace}} implemented through \texttt{check\_outbound\_replace}. In this procedure, all those PoIs that have been suspended would be replaced by other PoIs which are not suspended. The information about these PoIs would come from ``friends'' (having strong ties) of a vehicle. The whole suspended row in plan matrix of the vehicle would be replaced by a not-suspended row attained from a friend. Algorithm \ref{alg:replace} provides pseudocode of the procedure \textit {replace}. After replacement, \texttt{check\_outbound\_Blacklist} procedure would be called to select a PoI. Replace has two varieties 

\begin{algorithm}
\scriptsize
\caption{Replace Algorithm}\label{alg:replace}
\begin{algorithmic}[2]
\State $i \gets $0
\While{($i < 3$)}
	\State ${t} \gets $plan[i,4]
	\If{$(s = false)$}
		\State $i \gets i++$            
	\Else
		\State $j \gets 0$
      	\State $c \gets length[vehiclesdata]$
		\While{($j < c$)}
        	\State $t \gets friend \ j \ in \ the \ list \ of \ friends \ with \ strong \ tie$
        	\State $fplan \gets plan \ of \ f $
        	\State $k \gets 0$
			\While{($k < 3$)}            
				\State $ss \gets fplan[k,4]$
				\If{$(ss = false)$}
					\State $plan[i] \gets fplan[k] $
					\State $k \gets 3$
					\State $j \gets c$
				\Else
					\State $k \gets k++$
				\EndIf
				\State $j \gets j++$
			\EndWhile
		\EndWhile
	\EndIf		
\EndWhile
\end{algorithmic}
\end{algorithm}

\begin {itemize}

\item \textit{Replacement without triadic closure:} Replacement of blacklisted PoIs can be performed without considering the internal social networking dynamics, as mentioned above.

\begin{algorithm}
\scriptsize
\caption{triadic closure}\label{alg:TC}
\begin{algorithmic}[3]
\State $nof \gets number \ of \ friends \ of \ this \ vehicle$
\If{$(nof > 1)$} 
\State $fv \gets random[0-nof-1]$
\State $sv \gets random[0-nof-1]$
\If{$(fv != sv)$} 
\If {$fv \ has \ strong \ tie \ with \ this \ vehicle $}
\If {$sv \ is \ not \ friend \ of \ fv$} 
\State $make \ sv \ friend \ of \ fv$ 
\State $create \ strong \ tie \ between \ two \ vehicle$
\EndIf
\If {$fv \ is \ not \ friend \ of \ sv}$
\State $make \ fv \ friend \ of \ sv$
\State $create \ strong \ tie \ between \ two$
\EndIf 
\EndIf
\EndIf
\EndIf
\end{algorithmic}
\end{algorithm}
\item \textit{Replacement with triadic closure:} Replacement of blacklisted PoIs can also be performed while considering the internal social networking dynamics. One of the most basic and most important dynamics of social networks evolution is ``triadic closure''. The triadic closure property states that, if a node \textit{a} in a social network has strong ties with two nodes \textit{b} and \textit{c}, then, there is a likelihood that \textit{b} and \textit{c} would becomes friends themselves some times in the future. Algorithm \ref{alg:TC} provides pseudocode performing triadic closure just before application of Algorithm \ref{alg:replace}.

\end {itemize}

\section{Simulation and Results} \label {sec:sim}

The model is created in NetLogo \cite{wilensky1999netlogo}, a multi-agent programming environment. A grid of cells of size $75 \times 75$ is used to populate agents. There are three types of agents, two static and one mobile. The static PoI and Home agents and randomly placed in the environment. The count of PoIs is 15, available for 500 homes. Vehicles are mobile agents which are $10\%$ more than homes; thus, ensuring attachment of more than one vehicles to a home. A vehicles are attached randomly to a home. The users (owners of the vehicles) are abstract entities encapsulated within vehicles. 

The basic purpose of the vehicles is to visit PoIs as per plan at their prescribed time. The three strategies adopted (as explained before) are evaluated based on three parameters:
\begin {itemize}
\item quality-index: When a vehicle visits a PoI, it experiences the quality offered by that PoI at that time. In one week, a vehicle has to visit three PoIs, that is why, the quality-index is the average of three experiences of the vehicle at each hour. The quality-index of the whole simulation space is then an aggregation of all these experiences divided by vehicles count. 
\item connectivity-index: When two vehicles are at some PoI at the same time, they connect with each other, irrespective of type of tie they may have (weak or strong). The connectivity-index of the whole simulation space is an aggregation of all current (at this hour) connections divided by vehicles count.   
\item PoI-utilization: The population of vehicles at a PoI at an hour determines its current utilization. Derived from it is the standard deviation of utilization of all PoIs.
\end {itemize}

\begin{figure}
  \includegraphics[width=\textwidth]{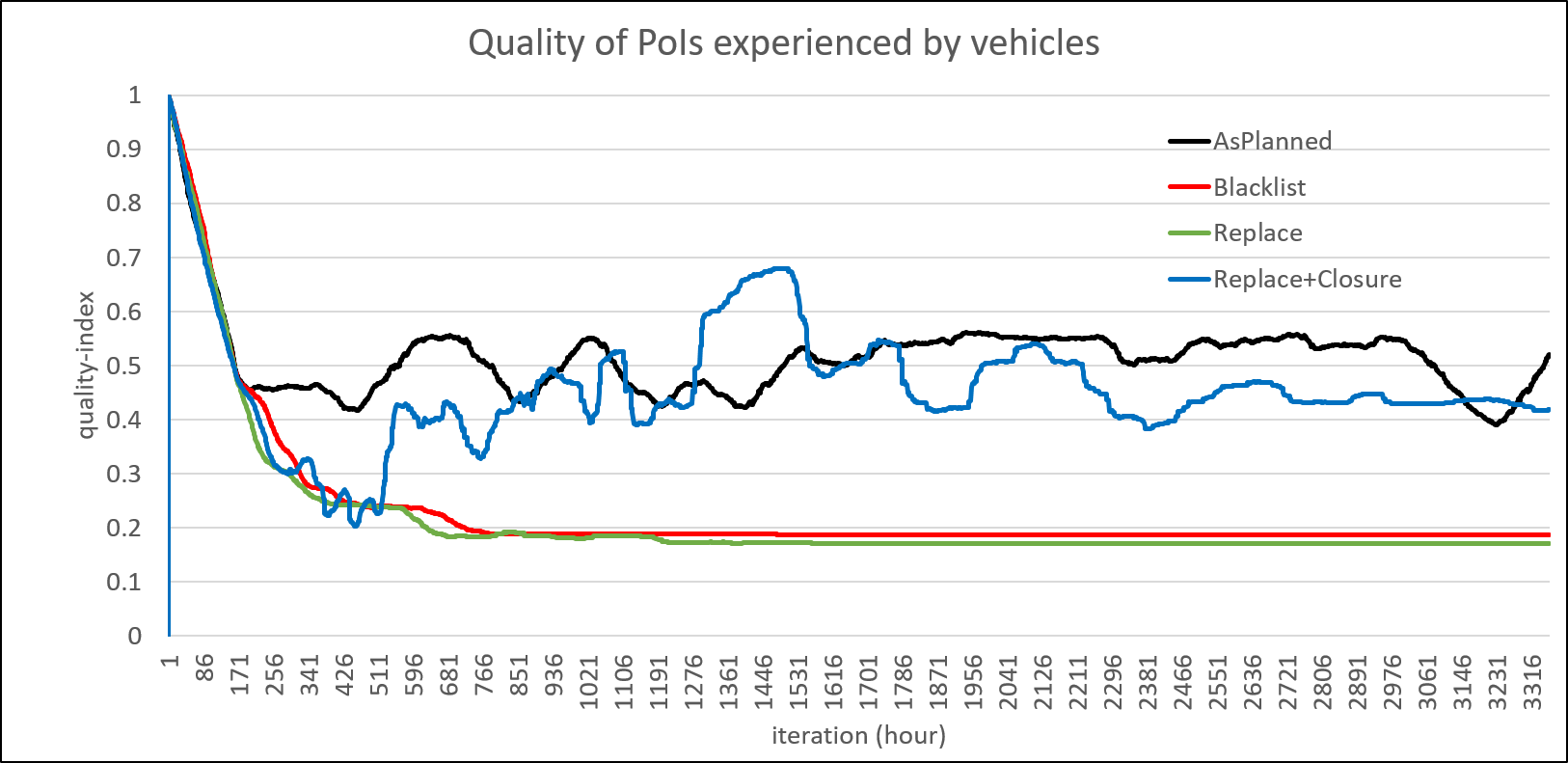}
  \caption{Comparison of Strategies ($th_{strong-tie} = 5$): Average quality of PoIs experienced by vehicles averaged across 100 simulation runs.}
  \label{fig:qg}
\end{figure}

The graph shown in Figure \ref{fig:qg} presents time series of average quality of PoIs experienced by all the vehicles for a span equal to twenty weeks and averaged across 100 simulation runs. A comparison between four strategies reveals the following:

\begin{itemize}

\item Blacklist: Once a PoI is blacklisted, it is not visited by a vehicle anymore. Less number of visits are performed hence reducing the average quality-index, due to the reason that after the first week, only those PoIs are visited which satisfy users / vehicles expectation.   

\item AsPlanned: Although, the above strategy (Blacklist) presents an extreme case, it can be argued to be the case. But, continuously visiting the same PoIs, which is not up to user expectations is less likely to happen. However, due to randomness in quality parameter of PoIs, the PoIs that did not perform according to the expectations previously, may perform well this time. This is what exactly is revealed in the simulation results, in which, after the first week, quality-index fluctuates around the middle (0.5).     

\item Replace: The replace strategy replaces a blacklisted PoI that a vehicle wants to visit by a PoI recommended by a friend. Since a friend is a strong tie, the value of threshold (repeated encounters) is important. A higher value of threshold ($th_{strong-tie} = 5$), such as 5 in this case, would delay replacement process. In this case, the process is so delayed that no vehicle has new information for others as all PoIs are already blacklisted.     

\item Replace with Closure: The triadic closure provides novel information about uncommon PoIs. That is the reason, most of the time, this strategy competes with the AsPlanned strategy. 
 
\end{itemize}

\begin{figure}
  \includegraphics[width=\textwidth]{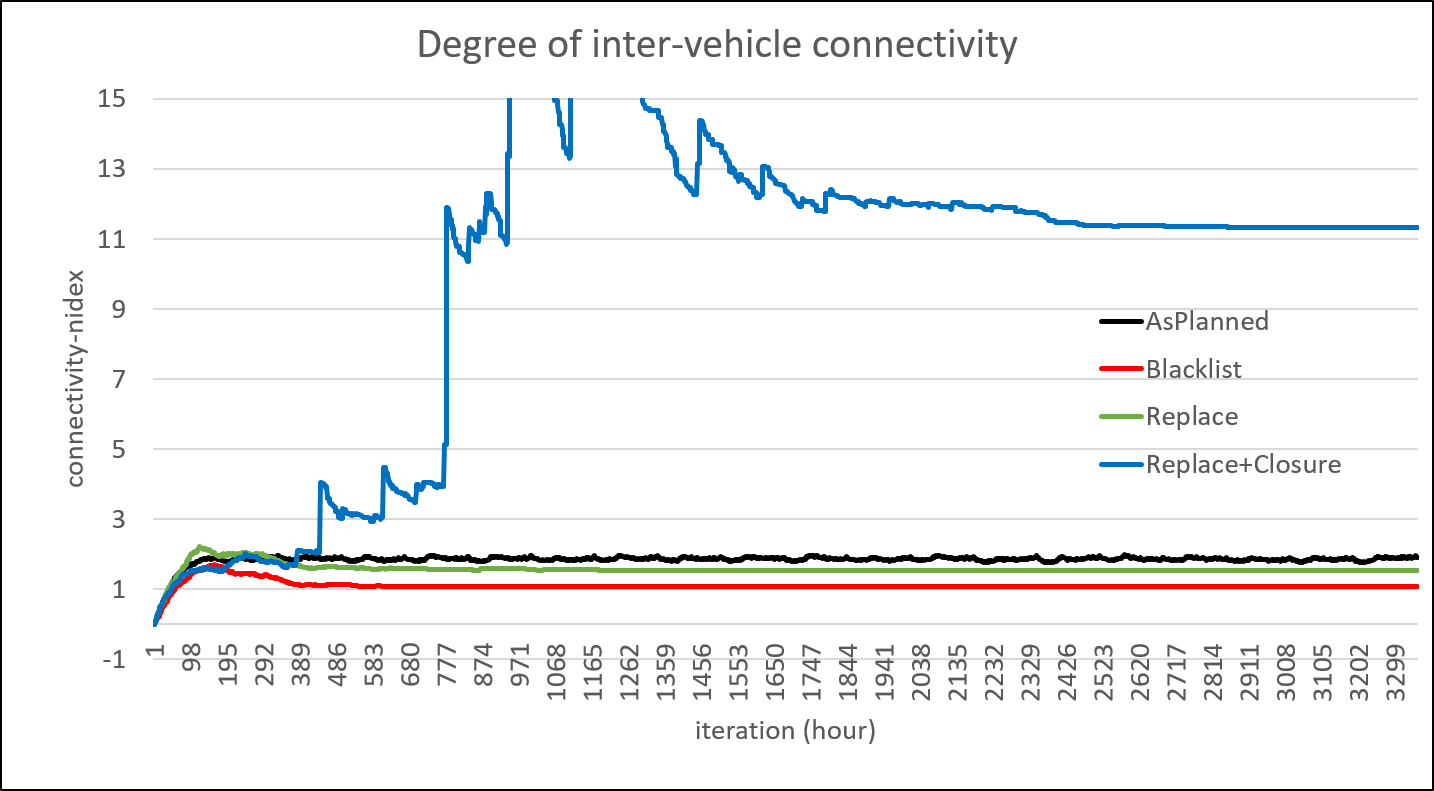}
  \caption{Comparison of Strategies ($th_{strong-tie} = 5$): Average connectivity of vehicles averaged across 100 simulation runs.}
  \label{fig:cg}
\end{figure}

The graph shown in Figure \ref {fig:cg} verifies the above findings relating the quality index with the average degree of vehicles. The connectivity of strategy replace with closure is by far higher than the other three strategies.

\begin{figure}
  \includegraphics[width=\textwidth]{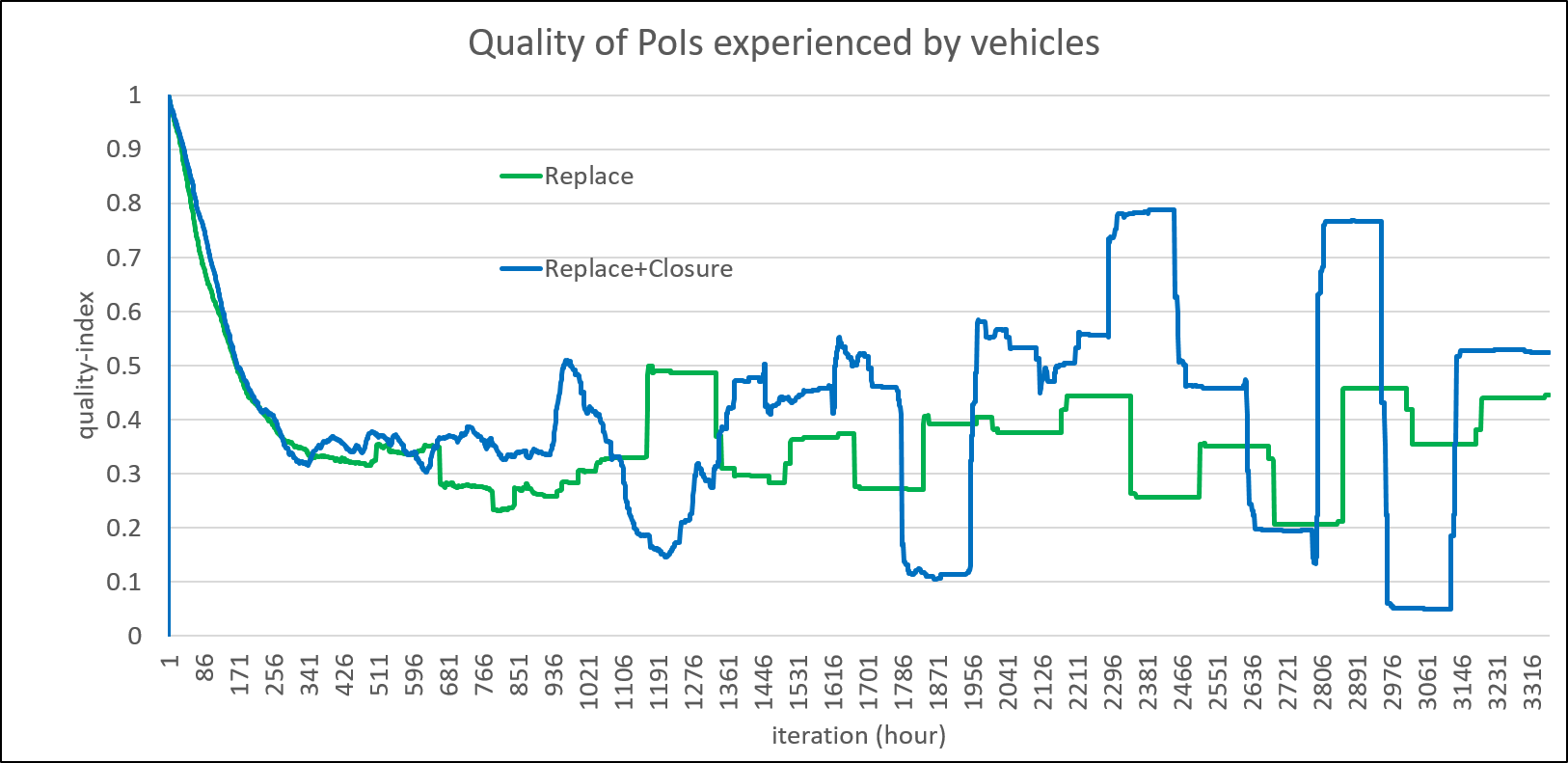}
  \caption{Comparison of Strategies ($th_{strong-tie} = 2$): Average quality of PoIs experienced by vehicles averaged across 100 simulation runs.}
  \label{fig:qg2}
\end{figure}

\begin{figure}
  \includegraphics[width=\textwidth]{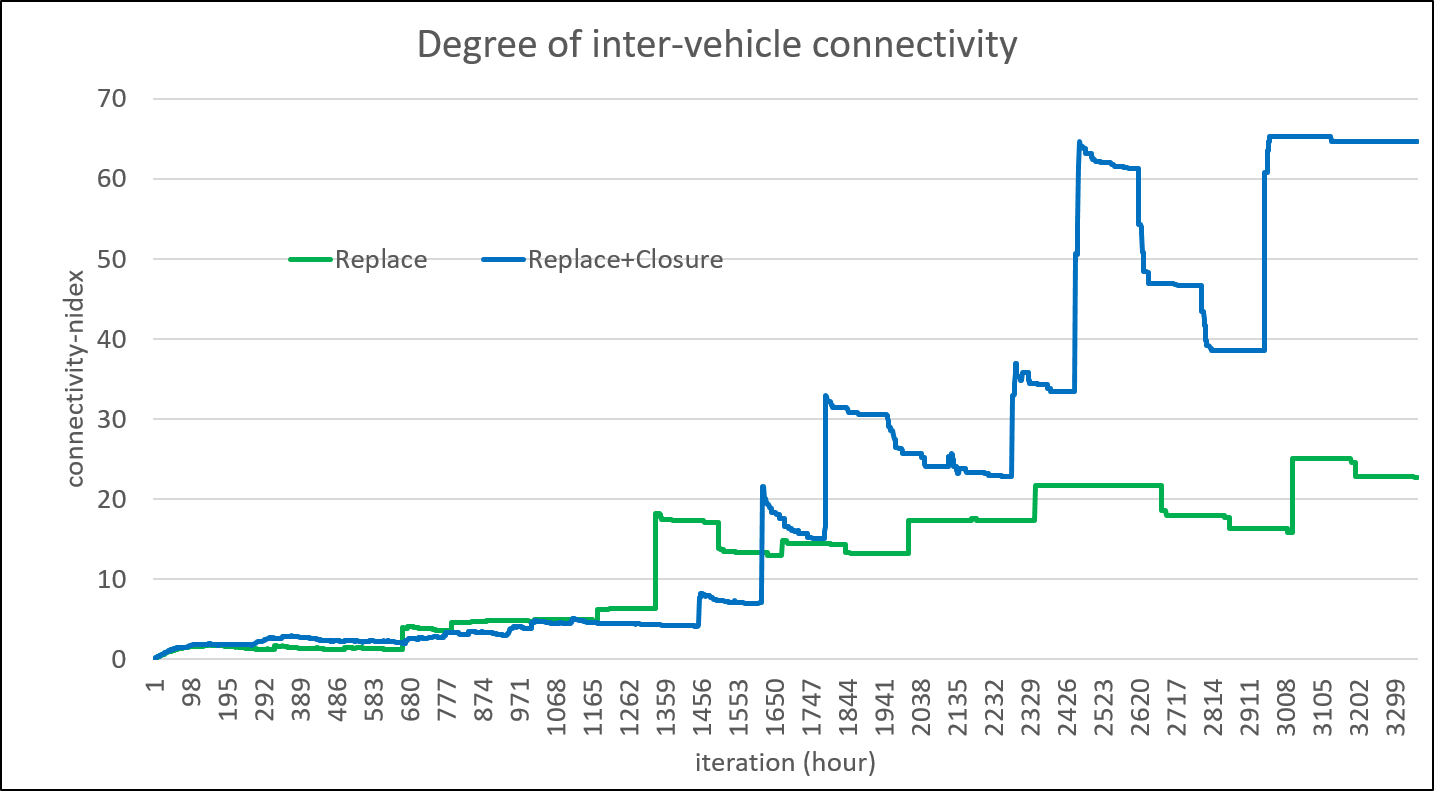}
  \caption{Comparison of Strategies ($th_{strong-tie} = 2$): Average connectivity of vehicles averaged across 100 simulation runs.}
  \label{fig:cg2}
\end{figure}

With decrease in encounter threshold of strong tie ($th_{strong-tie} = 2$), the Replace strategy starts competing with Replace strategy with triadic closure. It happens in patches through, where connectivity generally increases jumping up from one phase to another, consequently, increasing the quality index in an incremental way as well (See Figure \ref{fig:qg2} and Figure \ref{fig:cg2}). Overall, in terms of quality index replace and replace with triadic closure for $th_{strong-tie} = 2$ works much better than $th_{strong-tie} = 5$.

\begin{figure}
  \includegraphics[width=\textwidth]{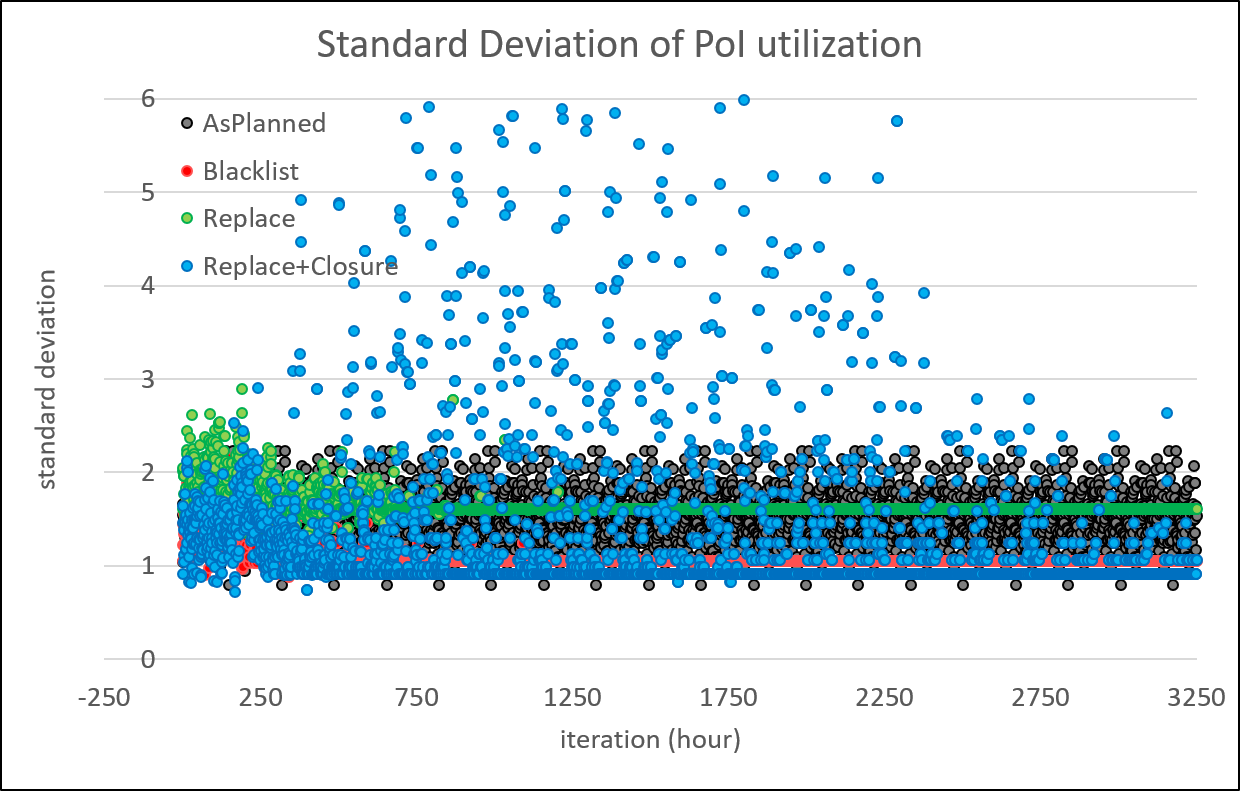}
  \caption{Comparison of Strategies ($th_{strong-tie} = 5$): Average standard deviation of PoIs utilization across 100 simulation runs.}
  \label{fig:sdg}
\end{figure}

\begin{figure}
 \includegraphics[width=\textwidth]{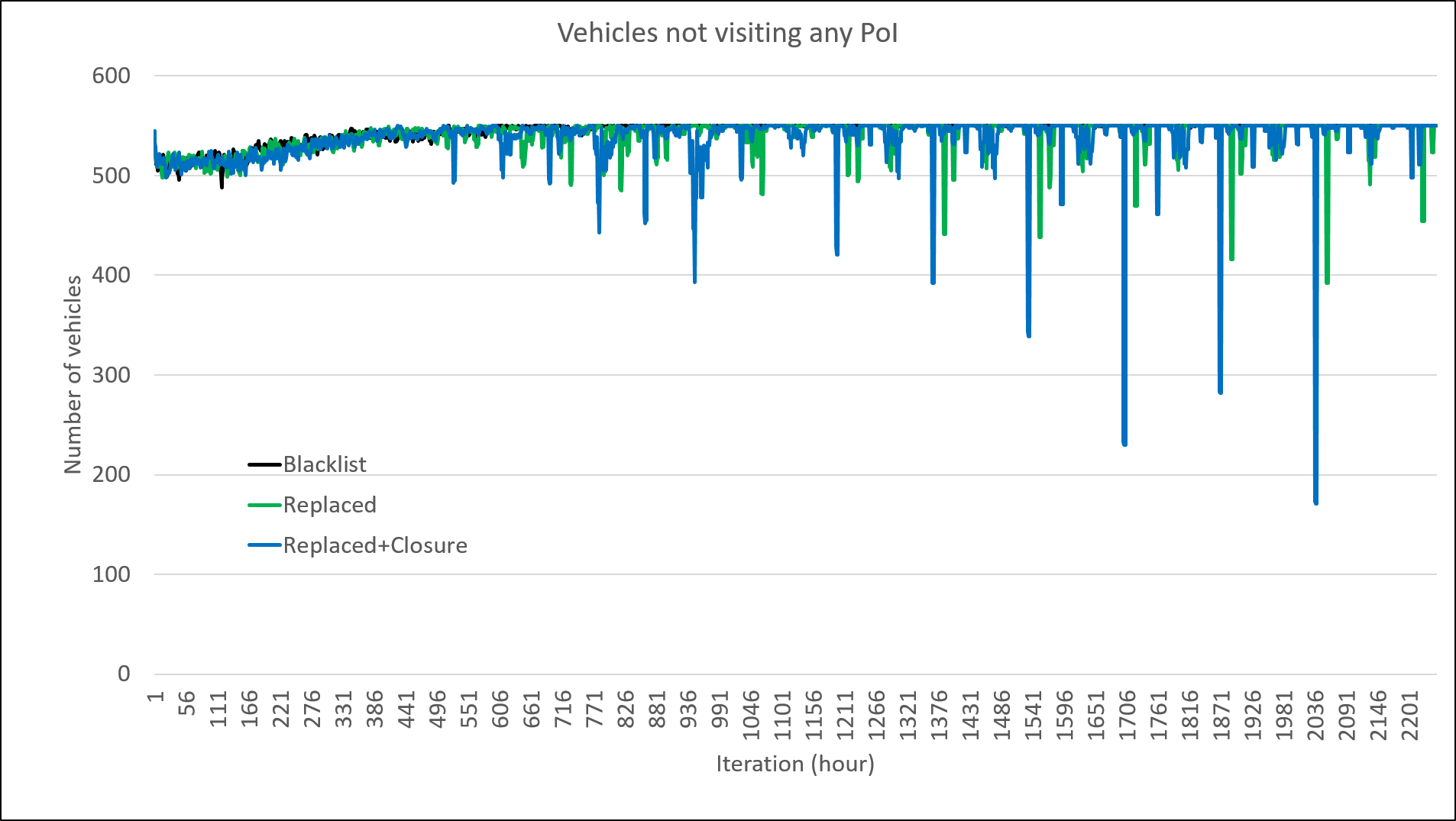}
  \caption{Comparison of Strategies ($th_{strong-tie} = 5$): Number of vehicles not visiting any PoI (one simulation run per strategy).}
  \label{fig:novisit}
\end{figure}

Another exciting aspect that is studies is utilization of PoIs. Although there is no safeguard against it in the model itself, it is undesirable to have high asymmetry in PoIs utilization, which is achieved through standard deviation of PoI utilization (SDU). The graph shown in Figure \ref {fig:sdg} compares four strategies. It can be seen that SDU of strategy blacklist is least because there are no visits after week 1, evidenced by the graph shown in Figure \ref{fig:novisit}. The strategy replace also has a low SDU after first few weeks, but this is also due to no visits. The strategy replace with closure has an erratic behavior, sometimes high and sometimes low. But contrary to the above, it is evident (from Figure \ref{fig:novisit}) that low SDU is not due to no visits, but, due to the symmetric utilization of PoIs which are not blacklisted. No doubt, high SDU is due to extreme asymmetry in PoIs utilization. 

\begin{figure}
  \includegraphics[width=\textwidth]{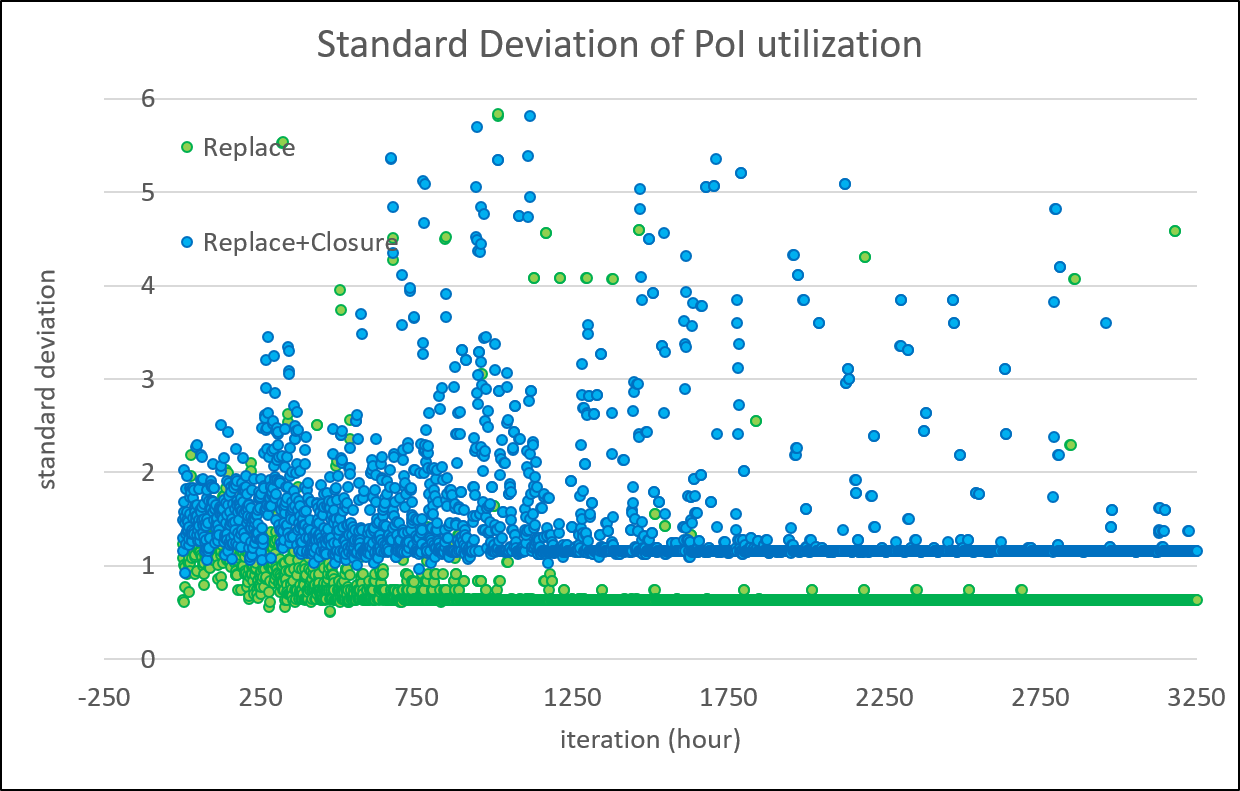}
  \caption{Comparison of Strategies ($th_{strong-tie} = 2$): Average standard deviation of PoIs utilization across 100 simulation runs.}
  \label{fig:sdg2}
\end{figure}

The inactivity of replace strategy is compensated by reducing from $th_{strong-tie} = 5$ to $th_{strong-tie} = 2$. Compare strategy replace with strategy replace with closure in the graph shown in Figure \ref{fig:sdg2}. Overall, again, in terms of PoIs utilization, replace and replace with triadic closure for $th_{strong-tie} = 2$ works much better than $th_{strong-tie} = 5$.

\section {Discussion and Conclusion} \label {sec:disc}

\begin{figure}
  \includegraphics[width=\textwidth]{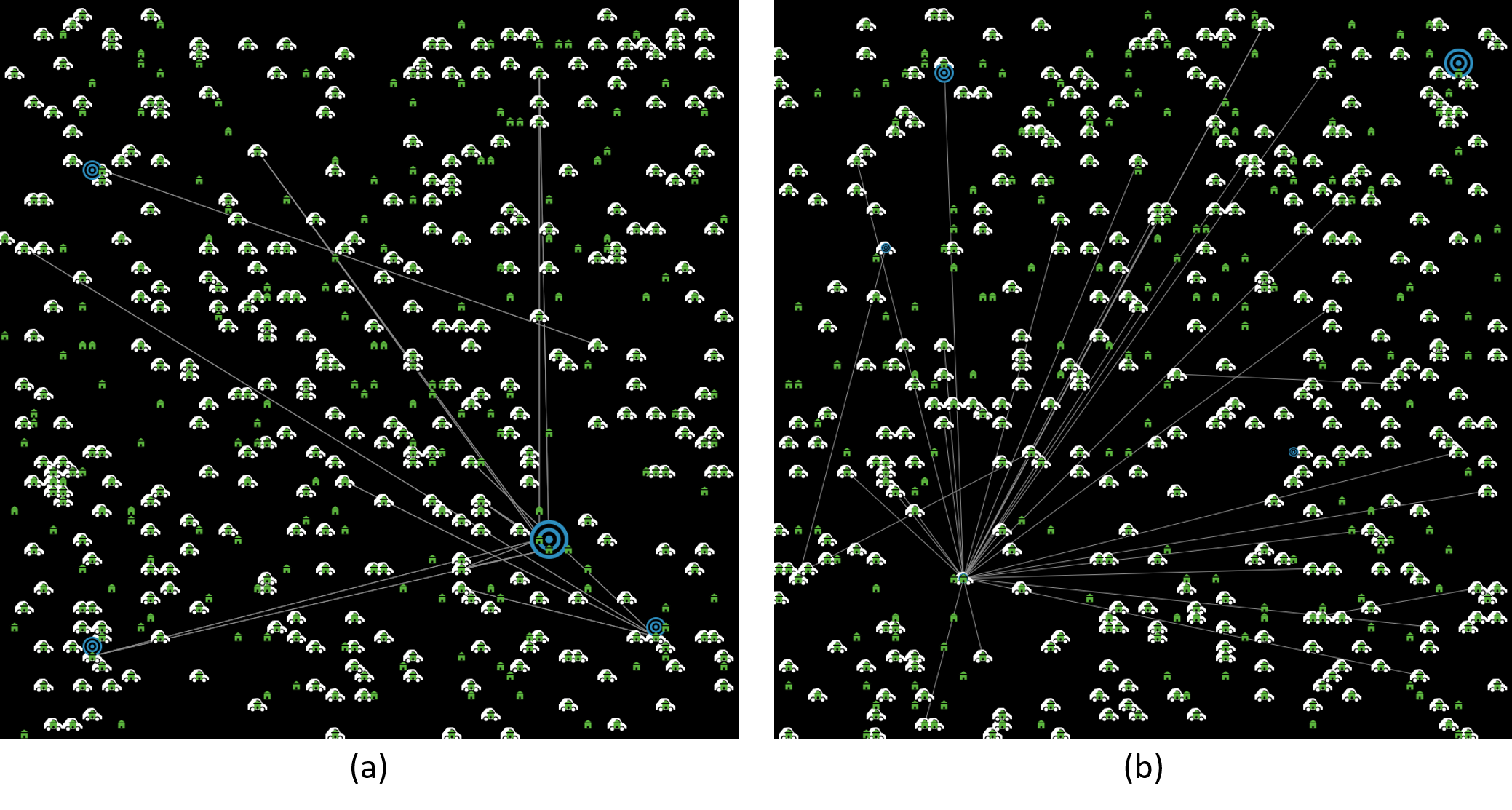}
  \caption{Comparison of Strategies ($th_{strong-tie} = 5$): Connectivity and PoI utilization at iteration 800. (a) Replace, (b) Replace with Closure.}
  \label{fig:5}
\end{figure}

\begin{figure}
  \includegraphics[width=\textwidth]{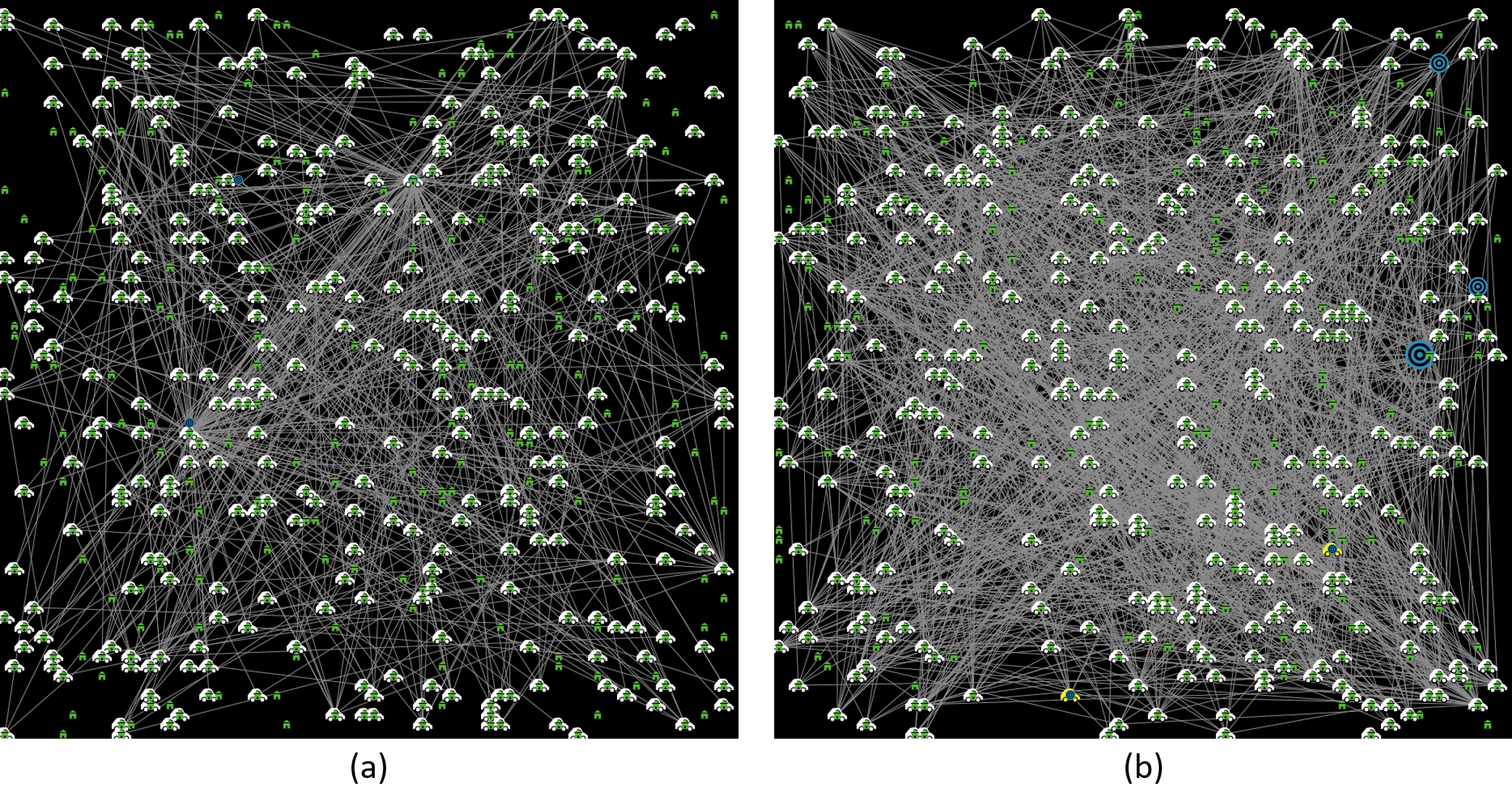}
  \caption{Comparison of Strategies ($th_{strong-tie} = 2$): Connectivity and PoI utilization at iteration 800. (a) Replace, (b) Replace with Closure.}
  \label{fig:2}
\end{figure}

In a dynamically changing quality of service provisioning by PoIs, the vehicles with blacklisted PoIs in their plan can utilize their strong ties to have novel information about some PoIs those are not blacklisted at a particular time. It is evident from connectivity comparison between replace and replace with closure in the simulation screenshots shown in Figure \ref{fig:5}, in which connectivity increases from (a) to (b). Increase in connectivity is responsible of increasing the quality index as depicted in Figure \ref{fig:qg} and \ref{fig:cg}. Same is true when threshold is decreased from $th_{strong-tie} = 5$ to $th_{strong-tie} = 2$ (see Figure \ref{fig:2} and corresponding graphs in Figure \ref{fig:qg2} and \ref{fig:cg2}). These results verify \textit{Hypothesis 1}.

However, a delay in the information sharing, particularly without triadic closure would not help. This is evident from inability of replace strategy to perform well if $th_{strong-tie} = 5$, whereas, it performs reasonably well if $th_{strong-tie} = 2$. This verifies \textit{Hypothesis 2}. In combination, the following can be concluded:

\noindent \textbf{``Novelty in information is directly proportional to how early the information sharing starts, however, triadic closure is capable of producing novel information even if information sharing starts late.''}  

In Figure \ref{fig:5} and \ref{fig:2}, the size of PoIs (blue circles) depicts their utilization. For $th_{strong-tie} = 5$ and $th_{strong-tie} = 2$ both, when comparing (a) with (b), the variation in sizes in less in (a) and more in (b). Hence, SDU is less in case of replace and more in replace with closure. It is also evident from Figure \ref{fig:sdg} and \ref {fig:sdg2}. A cross comparison among Figure \ref{fig:sdg} and \ref {fig:sdg2} reveals that for lesser threshold, SDU is even less. Moreover, at $th_{strong-tie} = 5$, replace with closure works better at the early stages of the simulation when compared with replace. This is opposite in case of $th_{strong-tie} = 2$, replace works better at the early stages of the simulation when compared with replace with closure. 

Overall, SDU is less in the early stages of the simulation and then it becomes erratic, fluctuating between extremely high values to extremely low values. Therefore, \textit{Hypothesis 3} is conditionally verified and can be stated as:

\noindent \textbf{``Fair distribution of PoIs is guaranteed at the start but not later.''}  

This brings us to the first outlook of the study: ``How can we guarantee a fair distribution of PoIs even in the later stages of the proceedings?'' The fact that blacklisted PoIs goes out of the system once and for all, and that they have no means to announce their updated quality even if it is high, it is natural to provide a mechanism of PoIs becoming part of the systems again. In real-life this can be related to the concept of memory and historical affiliation. Due to human ability to forget, and retain a long lasting relationships even after recent unpleasant experiences, a model extension incorporating these changes makes sense.

Another aspect of investigation in the future is finding a way to stop rapid fluctuations in SDU. The rapid fluctuations in SDU happen from herding of most vehicles to a few PoIs (resulting in extremely high values of SDU) to no activity (resulting in extremely low values of SDU) or vice versa. To average out these extreme cases, we intend to incorporate a cooperation strategy between two vehicles, alternating a PoI utilization between them, so that no PoI is loaded to its extreme.   

In conclusion, closure of social ties and its timing plays an important role in dispersion of novel information. As the network evolves as a result of incremental interactions, recommendations guaranteeing a fair distribution of vehicles across equally good competitors is not possible, particularly at the later stages of interactions.  
     
\section*{References}

\bibliography{main}

\begin{thebibliography}{10}
\expandafter\ifx\csname url\endcsname\relax
  \def\url#1{\texttt{#1}}\fi
\expandafter\ifx\csname urlprefix\endcsname\relax\def\urlprefix{URL }\fi
\expandafter\ifx\csname href\endcsname\relax
  \def\href#1#2{#2} \def\path#1{#1}\fi

\bibitem{weiser1991computer}
M.~Weiser, The computer for the 21st century, Scientific american 265~(3)
  (1991) 94--104.
\newblock \href {http://dx.doi.org/10.1145/329124.329126}
  {\path{doi:10.1145/329124.329126}}.

\bibitem{gubbi2013internet}
J.~Gubbi, R.~Buyya, S.~Marusic, M.~Palaniswami, Internet of things (iot): A
  vision, architectural elements, and future directions, Future generation
  computer systems 29~(7) (2013) 1645--1660.

\bibitem{wortmann2015internet}
F.~Wortmann, K.~Fl{\"u}chter, Internet of things, Business \& Information
  Systems Engineering 57~(3) (2015) 221--224.
\newblock \href {http://dx.doi.org/10.1007/s12599-015-0383-3}
  {\path{doi:10.1007/s12599-015-0383-3}}.

\bibitem{bakici2013smart}
T.~Bak{\i}c{\i}, E.~Almirall, J.~Wareham, A smart city initiative: the case of
  barcelona, Journal of the Knowledge Economy 4~(2) (2013) 135--148.
\newblock \href {http://dx.doi.org/10.1007/s13132-012-0084-9}
  {\path{doi:10.1007/s13132-012-0084-9}}.

\bibitem{neirotti2014current}
P.~Neirotti, A.~De~Marco, A.~C. Cagliano, G.~Mangano, F.~Scorrano, Current
  trends in smart city initiatives: Some stylised facts, Cities 38 (2014)
  25--36.
\newblock \href {http://dx.doi.org/10.1016/j.cities.2013.12.010}
  {\path{doi:10.1016/j.cities.2013.12.010}}.

\bibitem{evans2011internet}
D.~Evans, The internet of things: How the next evolution of the internet is
  changing everything, CISCO white paper 1~(2011) (2011) 1--11.

\bibitem{fangchun2014overview}
Y.~Fangchun, W.~Shangguang, L.~Jinglin, L.~Zhihan, S.~Qibo, An overview of
  internet of vehicles, China Communications 11~(10) (2014) 1--15.
\newblock \href {http://dx.doi.org/10.1109/CC.2014.6969789}
  {\path{doi:10.1109/CC.2014.6969789}}.

\bibitem{dandala2017internet}
T.~T. Dandala, V.~Krishnamurthy, R.~Alwan, Internet of vehicles (iov) for
  traffic management, in: Computer, Communication and Signal Processing
  (ICCCSP), 2017 International Conference on, IEEE, 2017, pp. 1--4.
\newblock \href {http://dx.doi.org/10.1109/ICCCSP.2017.7944096}
  {\path{doi:10.1109/ICCCSP.2017.7944096}}.

\bibitem{alam2015toward}
K.~M. Alam, M.~Saini, A.~El~Saddik, Toward social internet of vehicles:
  Concept, architecture, and applications, IEEE Access 3 (2015) 343--357.
\newblock \href {http://dx.doi.org/10.1109/ACCESS.2015.2416657}
  {\path{doi:10.1109/ACCESS.2015.2416657}}.

\bibitem{cunha2016data}
F.~Cunha, L.~Villas, A.~Boukerche, G.~Maia, A.~Viana, R.~A. Mini, A.~A.
  Loureiro, Data communication in vanets: Protocols, applications and
  challenges, Ad Hoc Networks 44 (2016) 90--103.
\newblock \href {http://dx.doi.org/10.1016/j.adhoc.2016.02.017}
  {\path{doi:10.1016/j.adhoc.2016.02.017}}.

\bibitem{abbani2011managing}
N.~Abbani, M.~Jomaa, T.~Tarhini, H.~Artail, W.~El-Hajj, Managing social
  networks in vehicular networks using trust rules, in: Wireless Technology and
  Applications (ISWTA), 2011 IEEE Symposium on, IEEE, 2011, pp. 168--173.

\bibitem{atzori2012social}
L.~Atzori, A.~Iera, G.~Morabito, M.~Nitti, The social internet of things
  (siot)--when social networks meet the internet of things: Concept,
  architecture and network characterization, Computer networks 56~(16) (2012)
  3594--3608.
\newblock \href {http://dx.doi.org/10.1016/j.comnet.2012.07.010}
  {\path{doi:10.1016/j.comnet.2012.07.010}}.

\bibitem{atzori2015social}
L.~Atzori, A.~Iera, G.~Morabito, Social internet of things: turning smart
  objects into social objects to boost the iot, Newsletter.

\bibitem{atzori2014smart}
L.~Atzori, A.~Iera, G.~Morabito, From" smart objects" to" social objects": The
  next evolutionary step of the internet of things, IEEE Communications
  Magazine 52~(1) (2014) 97--105.
\newblock \href {http://dx.doi.org/10.1109/MCOM.2014.6710070}
  {\path{doi:10.1109/MCOM.2014.6710070}}.

\bibitem{lee2015cyber}
J.~Lee, B.~Bagheri, H.-A. Kao, A cyber-physical systems architecture for
  industry 4.0-based manufacturing systems, Manufacturing Letters 3 (2015)
  18--23.
\newblock \href {http://dx.doi.org/10.1016/j.mfglet.2014.12.001}
  {\path{doi:10.1016/j.mfglet.2014.12.001}}.

\bibitem{festag2014cooperative}
A.~Festag, Cooperative intelligent transport systems standards in europe, IEEE
  communications magazine 52~(12) (2014) 166--172.
\newblock \href {http://dx.doi.org/10.1109/MCOM.2014.6979970}
  {\path{doi:10.1109/MCOM.2014.6979970}}.

\bibitem{smaldone2008roadspeak}
S.~Smaldone, L.~Han, P.~Shankar, L.~Iftode, Roadspeak: enabling voice chat on
  roadways using vehicular social networks, in: Proceedings of the 1st Workshop
  on Social Network Systems, ACM, 2008, pp. 43--48.
\newblock \href {http://dx.doi.org/10.1145/1435497.1435505}
  {\path{doi:10.1145/1435497.1435505}}.

\bibitem{Lei2016}
T.~Lei, S.~Wang, J.~Li, F.~Yang, A Cooperative Route Choice Approach via
  Virtual Vehicle in Internet of Vehicles, Springer International Publishing,
  Cham, 2016, pp. 194--205.
\newblock \href {http://dx.doi.org/10.1007/978-3-319-51969-2_16}
  {\path{doi:10.1007/978-3-319-51969-2_16}}.

\bibitem{dias2014cooperation}
J.~A. Dias, J.~J. Rodrigues, L.~Zhou, Cooperation advances on vehicular
  communications: A survey, Vehicular communications 1~(1) (2014) 22--32.

\bibitem{hu2013social}
X.~Hu, V.~Leung, K.~G. Li, E.~Kong, H.~Zhang, N.~S. Surendrakumar,
  P.~TalebiFard, Social drive: a crowdsourcing-based vehicular social
  networking system for green transportation, in: Proceedings of the third ACM
  international symposium on Design and analysis of intelligent vehicular
  networks and applications, ACM, 2013, pp. 85--92.

\bibitem{froehlich2009ubigreen}
J.~Froehlich, T.~Dillahunt, P.~Klasnja, J.~Mankoff, S.~Consolvo, B.~Harrison,
  J.~A. Landay, Ubigreen: investigating a mobile tool for tracking and
  supporting green transportation habits, in: Proceedings of the SIGCHI
  Conference on Human Factors in Computing Systems, ACM, 2009, pp. 1043--1052.

\bibitem{saremi2016participatory}
F.~Saremi, Participatory sensing fuel-efficient navigation system greengps,
  Ph.D. thesis, University of Illinois at Urbana-Champaign (2016).

\bibitem{smaldone2011cyber}
S.~Smaldone, C.~Tonde, V.~K. Ananthanarayanan, A.~Elgammal, L.~Iftode, The
  cyber-physical bike: A step towards safer green transportation, in:
  Proceedings of the 12th Workshop on Mobile Computing Systems and
  Applications, ACM, 2011, pp. 56--61.

\bibitem{sun2016internet}
Y.~Sun, H.~Song, A.~J. Jara, R.~Bie, Internet of things and big data analytics
  for smart and connected communities, IEEE Access 4 (2016) 766--773.
\newblock \href {http://dx.doi.org/10.1109/ACCESS.2016.2529723}
  {\path{doi:10.1109/ACCESS.2016.2529723}}.

\bibitem{zakir2015big}
J.~Zakir, T.~Seymour, K.~Berg, Big data analytics., Issues in Information
  Systems 16~(2).

\bibitem{park2014driver}
E.~Park, K.~J. Kim, Driver acceptance of car navigation systems: integration of
  locational accuracy, processing speed, and service and display quality with
  technology acceptance model, Personal and ubiquitous computing 18~(3) (2014)
  503--513.
\newblock \href {http://dx.doi.org/10.1007/s00779-013-0670-2}
  {\path{doi:10.1007/s00779-013-0670-2}}.

\bibitem{bao2015recommendations}
J.~Bao, Y.~Zheng, D.~Wilkie, M.~Mokbel, Recommendations in location-based
  social networks: a survey, GeoInformatica 19~(3) (2015) 525--565.
\newblock \href {http://dx.doi.org/10.1007/s10707-014-0220-8}
  {\path{doi:10.1007/s10707-014-0220-8}}.

\bibitem{zhang2017carstream}
M.~Zhang, T.~Wo, T.~Xie, X.~Lin, Y.~Liu, Carstream: an industrial system of big
  data processing for internet-of-vehicles, Proceedings of the VLDB Endowment
  10~(12) (2017) 1766--1777.

\bibitem{fagnant2014travel}
D.~J. Fagnant, K.~M. Kockelman, The travel and environmental implications of
  shared autonomous vehicles, using agent-based model scenarios, Transportation
  Research Part C: Emerging Technologies 40 (2014) 1--13.

\bibitem{easley2010networks}
D.~Easley, J.~Kleinberg, Networks, crowds, and markets: Reasoning about a
  highly connected world, Cambridge University Press, 2010.

\bibitem{wilensky1999netlogo}
U.~Wilensky, Netlogo.

\end{thebibliography}

\end{document}